# High Open-circuit Voltage of Graphene-Based Photovoltaic Cells Modulated by Layer-by-Layer Transfer

Kyuwook Ihm[†], Kyung-jae Lee[‡], Jong Tae Lim[§,*], Jae Wook Kwon[§], Tai-Hee Kang[†], Sukmin Chung[‡], Sukang Bae[||], Jin Ho Kim[||], Byung Hee Hong[||,*], Geun Young Yeom[§,*]

[†]*Pohang Accelerator Laboratory, Pohang, Kyungbuk 790-784, Korea*

[‡]*Department of physics, POSTECH, Pohang, Kyungbuk 790-784, Korea*

[§] *School of Advanced Materials Science and Engineering, Sungkyunkwan University, Suwon 440-746, Korea*

[||] *Department of Chemistry and SKKU advanced Institute of Nanotechnology, Suwon 440-746, Korea*

*To whom correspondence should be addressed. E-mail: (J. T. L.) lapbbi@skku.edu; (B. H. H.) byunghee@skku.edu; (G. Y. Y.): gyyeom@skku.edu

## ABSTRACT

Graphene has shown great application opportunities in future nanoelectronic devices due to its outstanding electronic properties. Moreover, its impressive optical properties have been attracting the interest of researchers, and, recently, the photovoltaic effects of a heterojunction structure embedded with few layer graphene (FLG) have been demonstrated. Here, we report the photovoltaic response of graphene-semiconductor junctions and the controlled open-circuit voltage ($V_{oc}$) with varying numbers of graphene layers. After unavoidably adsorbed contaminants were removed from the FLGs, by means of in situ annealing, prepared by layer-by-layer transfer of the chemically grown graphene layer, the work functions of FLGs showed a sequential increase as the graphene layers increase, despite of random interlayer-stacking, resulting in the modulation of photovoltaic behaviors of FLGs/Si interfaces. The surface photovoltaic effects observed here show an electronic realignment in the depth direction in the FLG heterojunction systems, indicating future potential toward solar devices utilizing the excellent transparency and flexibility of FLG.

Graphene has shown remarkable photonic properties, such as high transparency and wide absorption spectral range, as well as outstanding electronic and mechanical properties, so extensive interest has developed toward applying it to optoelectronic devices [1-8]. The success in growing a qualified wide-area few layer graphene (FLG), comparable to one exfoliated from graphite, using chemical vapor deposition method (CVD) enables researchers to readily apply it to optoelectronic devices and quantitatively analyze its driving mechanism [9]. The graphene heterojunction system is the most commonly encountered structure in the study of graphene itself and in building application-oriented interfaces. In these systems, graphene-substrate interactions induce significant effects on the observed results due to the atomic-order thinness of the graphene; the graphene surface is never free from the influence of interface effects. One of these effects is the surface photovoltage which can be found in the low coverage metal/semiconductor junction [10]. A similar effect was reported in the field effect transistor consisting of graphene on an insulating surface with metal pattern which showed a photo response driven by lateral band transition when the graphene crosses the different substrate surfaces, i.e., the edge effect [5, 6, 11, 12]. It is known that the optical properties of FLG directly depend on the number of layers due to different transparencies and absorbance processes [5, 6]. However, it is not well studied why and how photo-responses depend on the modulated electronic structures of FLG with different numbers of layers when it is applied to a heterojunction system. Recently, it had been reported that the work function of double layer graphene exfoliated from graphite differs from that of single layer graphene due to chemical stabilization induced by an interlayer interaction [13]. Here, we show that the work function of the FLG prepared by layer-by-layer transfer of the CVD-grown graphene increases as the number of layers increases, and this is directly connected to open-circuit voltage modulation of an FLG/Si photovoltaic cell.

The large-area graphene layers were synthesized by CVD of methane gas on Cu foils at 1000 ˚C [14]. After the graphene film was spin-coated at 3,000 rpm with 5 wt% polymethylmethacrylate (PMMA) in chlorobenzene, the underlying Cu foil was etched by 0.5 M aqueous $FeCl_3$ solution. Subsequently, the PMMA-supported graphene layer was transferred onto Si substrates and dried on a hot plate at 80 ˚C. Finally, acetone was used to remove the PMMA layer. Annealing and spectroscopic measurement of the samples were carried out at the 4B1 PES II Beamline at the Pohang Light Source in the Pohang Accelerator Laboratory [15]. The devices were mounted onto a grounded molybdenum holder, thermally linked by phosphor-boron nitride composites to a resistance heater. The top-lying FLGs in the devices were characterized via photoemission electron spectroscopy (PES) at several annealing steps in the analysis chamber (base pressure $4 \times 10^{-10}$ torr).

The Raman spectra showed that the transferred graphene film consists of >95% monolayers [14]. As the graphene layers were transferred one after another [16], the intensities of G and 2D band peaks increased together, but their ratios did not change significantly (Fig.1). This is because the hexagonal lattices of upper and lower layers are randomly oriented, unlike graphite, so that the original properties of each monolayer remain unchanged even after stacking into multilayers [17, 18]. This is clearly different from the case of multilayer graphene exfoliated from graphite crystals [19]. The Raman spectra of the graphene layers transferred onto Si don't show D peaks near 1300 $cm^{-1}$, indicating the high quality of the graphene films. The TEM image also shows that the graphene film is a high-quality monolayer without many defects (Fig. 1, inset. For more images, see figure S4 in ref. 14).

A schematic view of a prototype of the photovoltaic device is shown in Fig.2a. A graphene sheet (1ML, $10\times10$ $mm^2$) was repeatedly transferred onto a *p*-type Si (Density of $N_a$ (Boron) = $2\times 10^{15}$ $cm^{-2}$, $12\times12$ $mm^2 \times 525$ μm) surface until the intended number of layers was reached. On the corner of the graphene, an Al contact ($2\times2$ $mm^2$, 3000 Å) to collect electrons was formed by e-beam evaporation through a shadow mask. Figure 2b shows the time evolution of the open circuit voltage ($V_{oc}$) of the devices with 1 and 2 ML graphene layers as a photovoltaic response to a photon source (100 mW $cm^{-2}$). Initial fluctuations of $V_{oc}$ upon photon illumination rapidly disappeared, and it stabilized at values of 0.10 V for 1 ML and 0.17 V for 2 ML devices. This is a transient effect as the devices approach a thermal equilibrium. The current density versus the bias voltage of as prepared devices is plotted in Fig. 2c for 100 mW (AM 1.5G) illumination. The $V_{oc}$ of devices with 1, 2, 3, 4 graphene layers are -110, -190, -50 and -3 mV, respectively. Power conversion efficiencies ($\eta$) for devices with 1 and 2 ML graphene are about 0.01 % with a fill factor (FF) of about 0.23 for both with a short-current density ($J_{sc}$) of 0.22 and 0.42 mA cm$^{-2}$, respectively. However, the devices with 3 and 4 MLs had $\eta$ below 0.001 %.

The $V_{oc}$ is exclusively dependent on the work function difference between FLG and Si substrate as in the case of the metal/semiconductor (MS) structure. As the work functions of the Si substrates in all devices are almost the same, $V_{oc}$ is determined by the work function values of the FLGs. Except for 2 ML sample in Fig. 3c, $V_{oc}$ decreases as the number of graphene layers is increased. This is opposite to the result observed in Raman spectra, showing no significant difference between one and multilayer graphene. This implies that a weak interlayer interaction via $\pi$ orbitals is established despite of random stacking of graphene layers. To remove unexpected effects induced by contaminants adsorbed on the FLGs during chemical preparation processes, the samples were annealed at 500 °C for 12 hours in the vacuum chamber (base pressure: $4\times10^{-10}$ Torr) directly connected to the photoemission spectroscopy chamber. The *J-V* curves of the devices after annealing are shown in Fig. 3d. $V_{oc}$ decreased in sequence, -120, -50, -30 and -5 mV, as the number of graphene layers increased. This indicates that the work function difference of FLG from Si is decreased as the number of graphene layers increases. In *J-V* measurement electrons are collected by the aluminum contact through the devices, implying that the work function of FLG has to be lower than that of the Si substrate and that the work function of FLG is increased as the number of graphene layers increases. This result is consistent with the recent result of Yu et al. They showed the work function of the double layer graphene exfoliated from graphite was higher ($\Delta\phi$ = +0.12 eV) than that of single layer graphene ($\phi$ = 4.57 ± 0.05 eV) using scanning Kelvin probe microscopy and explained this as a result of chemical stabilization driven by interlayer interactions [13]. Figure 3a shows the onset of the secondary cut-off (SC), $E_{SC}$, of photoemission electron spectrum from as-introduced samples at a bias of -5 V using an ultraviolet source (He I: $h\nu$ = 21.2 eV). The variation of SC value is directly dependent on the change of work function, as $\Delta\phi = \Delta\{h\nu - (E_f - E_{SC})\} = \Delta E_{SC}$. Here, $E_f$ is the Fermi level of a sample. The inset of Fig. 3a shows the work function difference between as-introduced FLG, $\Delta\phi_G$, and the Si substrate ($\phi_{si}$ =

4.61 ± 0.02 eV). The work function of Si substrate beneath the FLG was estimated from the SC (gray line in Fig 3a) of the exposed Si surface by Ar+ ion sputtering until the signals of foreign atoms disappeared from the x-ray photoemission electron spectrum. The work function variation was similar with that of $V_{oc}$ in Fig. 2c, as the numbers of layers changed. Unexpectedly, 2 ML graphene had a higher work function and $V_{oc}$ than the others. Figure 3b shows the SC of FLG after annealing at 500 °C for 12 hours in the vacuum chamber. Interestingly, the SC value is well behaved, i.e., it increased as the number of graphene layers increased without any exception. This implies a sequential decrease of contact potential energy, $\Delta\phi_G$, which agrees well with the behavior of $V_{oc}$ in Fig. 2d. Note that the work function difference (0.04 eV) between 1 and 2 ML graphene is smaller than that (0.12 eV) of exfoliated ones [13]. This shows weak interlayer interactions of FLGs prepared by layer-by-layer transfer due to random interlayer stacking. The prominent dependence of $V_{oc}$ on $\Delta\phi_G$ indicates that the photovoltaic effects in the FLG/Si structure can be described by the photovoltaic effect of a MS structure. The MS cell's open-circuit voltage, $V_{OC} = n[\phi_B / q + (kT/q)\ln(I_s / A_c A^+ T^2)]$, mainly depends on the contact barrier, $\phi_B = \phi_{Si} - \phi_G = \Delta\phi_G$, and diode quality factor, $n$, because $kT$ at room temperature (~ 0.0259 eV) is far smaller than $\phi_B$. $A_c$ and $A^+$ are the contact area of the diode and the Richardson constant, respectively [20, 21]. The schematic view of electronic structure of FLG/Si structure is shown in Fig. 3c when $\phi_G < \phi_{Si}$ and Fig. 3d when $\phi_G > \phi_{Si}$. If the work function of FLG is smaller than that of Si substrate, Schottky contact is established with transition region spread into the Si substrate, where charge separation of the electron-hole pairs created by photons occurs. As the work function of the FLG increases with the number of graphene layer, the contact potential $\Delta\phi_G$ as well as the rectifying character responsible for the photovoltaic effect is decreased. This well explains why a negligible photovoltaic effect occurs in the samples with 3 and 4 ML graphene. The opposite case of Schottky contact is Ohmic contact when the work function of FLG exceeds that of Si as shown in Fig. 3d. In this case the majority carrier current passes easily in either direction because of low contact resistance.

In all devices $V_{oc}$ is about 100 mV smaller than that of the estimated Schottky barrier, $\Delta\phi_G$. If the FLG is approximated as a metal, the depletion width in the transition region of Si is about 0.56 μm from the equation $W \approx \sqrt{2\varepsilon\Delta\phi_G / qN_a}$, where $N_a$ is the density of boron. This indicates that the largest portion of the Si substrate acts as a series resistor in the devices. This explains the lower $V_{oc}$ than expected value.

The modification of FLG after annealing was characterized in the connected analysis chamber using photoemission electron spectroscopy. Figure 4 shows the C 1s spectra at $h\nu$ = 350 eV from synchrotron radiation source. The parameters of decomposed peaks are characterized by two divided parts: The parameters of the peaks originating from FLG include a Doniac-Sunjic line shape, using a Lorentzian line width of 178 meV and an asymmetry factor of 0.07 following the previous result [22], while the parameters of peaks of contaminant carbon species with asymmetric character were excluded due to weak coupling of contaminants with the graphene layer (See Supplementary Information, Tab. S1, for details of parameters). The C 1s spectrum of as-introduced 1 ML graphene was decomposed into one graphene peak and four peaks related with contaminants, which are integrated as the red area in the bottom spectra in Fig. 4. The considerable amount of uncontrollable contaminants explains the unexpected behavior appearing in the work function of as-introduced FLGs as a function of layers. However, the unexpectedly low work function of as-introduced 2 ML graphene, leading to high $V_{oc}$, shows another possible challenge to increase $V_{oc}$ by controlled doping of the graphene. Upon annealing the sample the C 1s spectra are well characterized by two peaks originating from FLG and another two peaks from contaminants. In the case of 1 ML graphene only one graphene related peak shown by blue area, named M, appears with two contaminant peaks named C1 and C2. In the 2 ML graphene, the monolayer peak, M, is decreased while a bulk graphene peak represented by black area is visible at 284.45 eV. In the 4 ML graphene, the monolayer peak is completely replaced by the bulk graphene peak B. This result is comparable to that of an exfoliated graphene sample whoes work function depends on the number of layers [13, 23]. The contaminant related peaks, C1 and C2, are gradually decreased and only a negligible amount is visible in 4 ML graphene. This indicates that the C1 and C2 contaminants are localized between the Si and FLG, where they are difficult to desorb by an annealing process.

In summary, we have shown the modulation of the work function of FLG by controlling the number of graphene layers grown by a CVD method. The photovoltaic response is found to directly depend on the number of graphene layers in the FLG/Si structure, similarly to the surface photovoltaic effects of a low coverage metal-semiconductor junction. The surface photovoltaic effect found here is a possible model system to estimate the modified electronic structure at the interface of a graphene heterojunction, which is commonly encountered in graphene research.

## FIGURE CAPTIONS

Figure 1. Raman spectra of graphene films transferred layer-by-layer onto a Si substrate. The 2D/G peak ratios don't change significantly as the number of layers increases from 1 to 4. The inset shows a TEM image of the monolayer graphene.

Figure 2. Photovoltaic responses of graphene/Si. (a) a schematic view of the sample. (b) Plot of open-circuit voltage versus time of 1 and 2 ML graphene/Si. Plot of J-V of as introduced samples with 1~4 ML graphene on Si, (c), and after annealing at 500 °C for 12 hour,(d).

Figure 3. Work function modulation of FLG and electronic structure at interface of FLG and Si. (a) Onset of secondary cut-off(SC) of UPS spectra of as introduced FLG (1~4 ML graphene) and the pristine Si surface. The work function difference between FLG and the pristine Si surface is shown in the inset. (b) Onset of SC of the FLG after annealing at 500 °C for 12 hour, with work function differences in the inset. (c) Electronic structure when Schottky contact is established. The electron-hole pairs photo-generated within the depletion region of Si are separated by the built-in potential. (d) Electronic structure, with Ohmic contact.

Figure 4. C 1s core level spectra taken at $h\nu$=350 eV. Peaks in blue and black area, named M and B, are the main features of monolayer graphene and bulk graphene, respectively. Both have a Doniac-Sunjic line shape [17]. The C 1s spectrum of as-introduced 1ML graphene, the bottom one, has one monolayer peak and four contaminant related peaks, which are integrated as the red area. C 1s spectra of 1~4 ML graphene after annealing at 500 °C are shown in the second to top figure. Two contaminant related peaks, shown by pink and green areas, decreased as the number of layers increased, while the monolayer peak, M, is replaced by the bulk peak, B, at 4 ML graphene.

## Supplementary information

Table S1. Best-fit results and estimated contribution to Doniac-Sunjun line shape, which is the characteristic feature of graphene [22]. Gaussian widths of all peaks are fixed at 752 meV.

Figure 1.

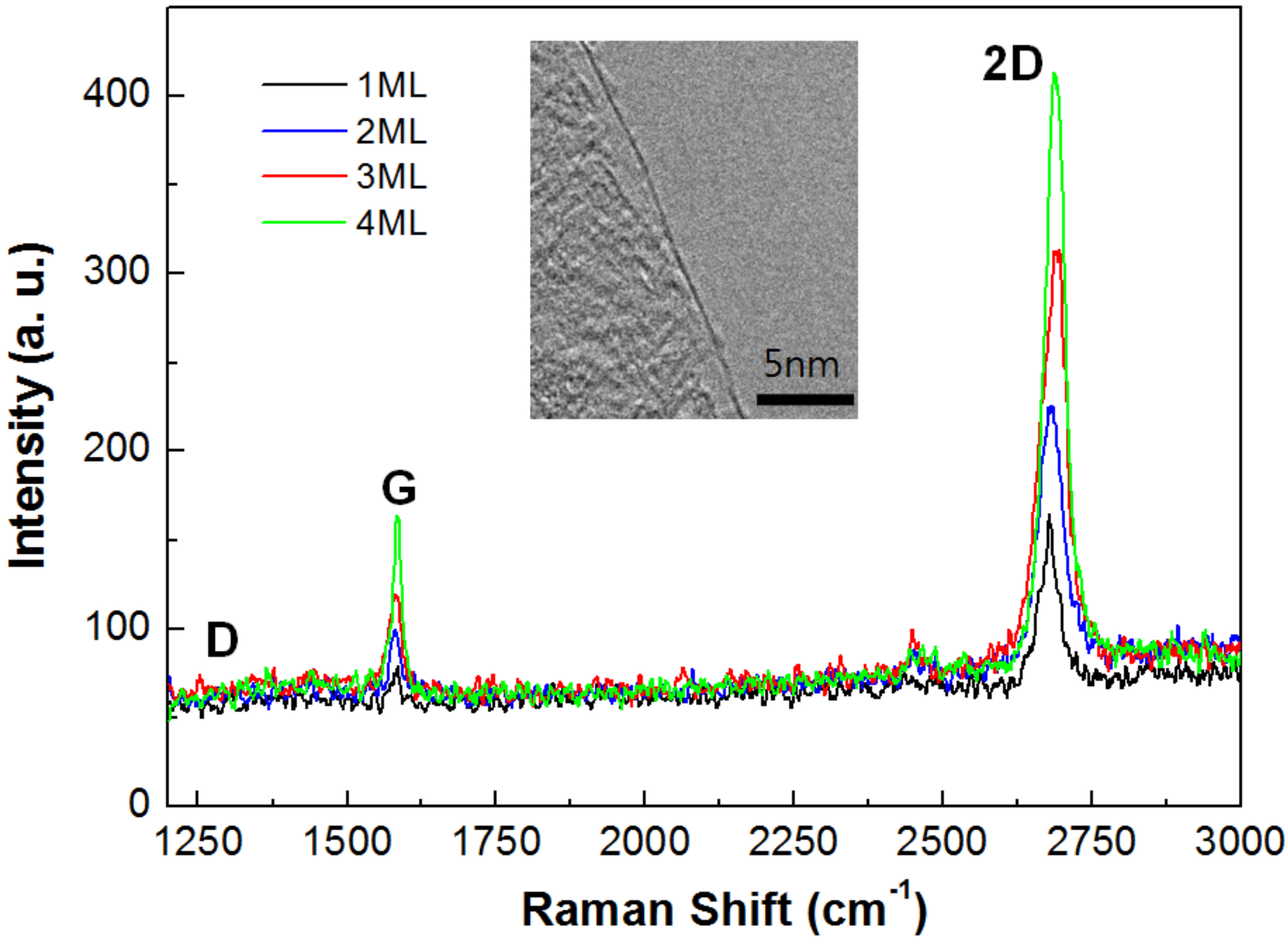

1ML
2ML
3ML
4ML
5nm
2D
G
D
Intensity (a. u.)
Raman Shift (cm$^{-1}$)
0
100
200
300
400
1250
1500
1750
2000
2250
2500
2750
3000

Figure 2.

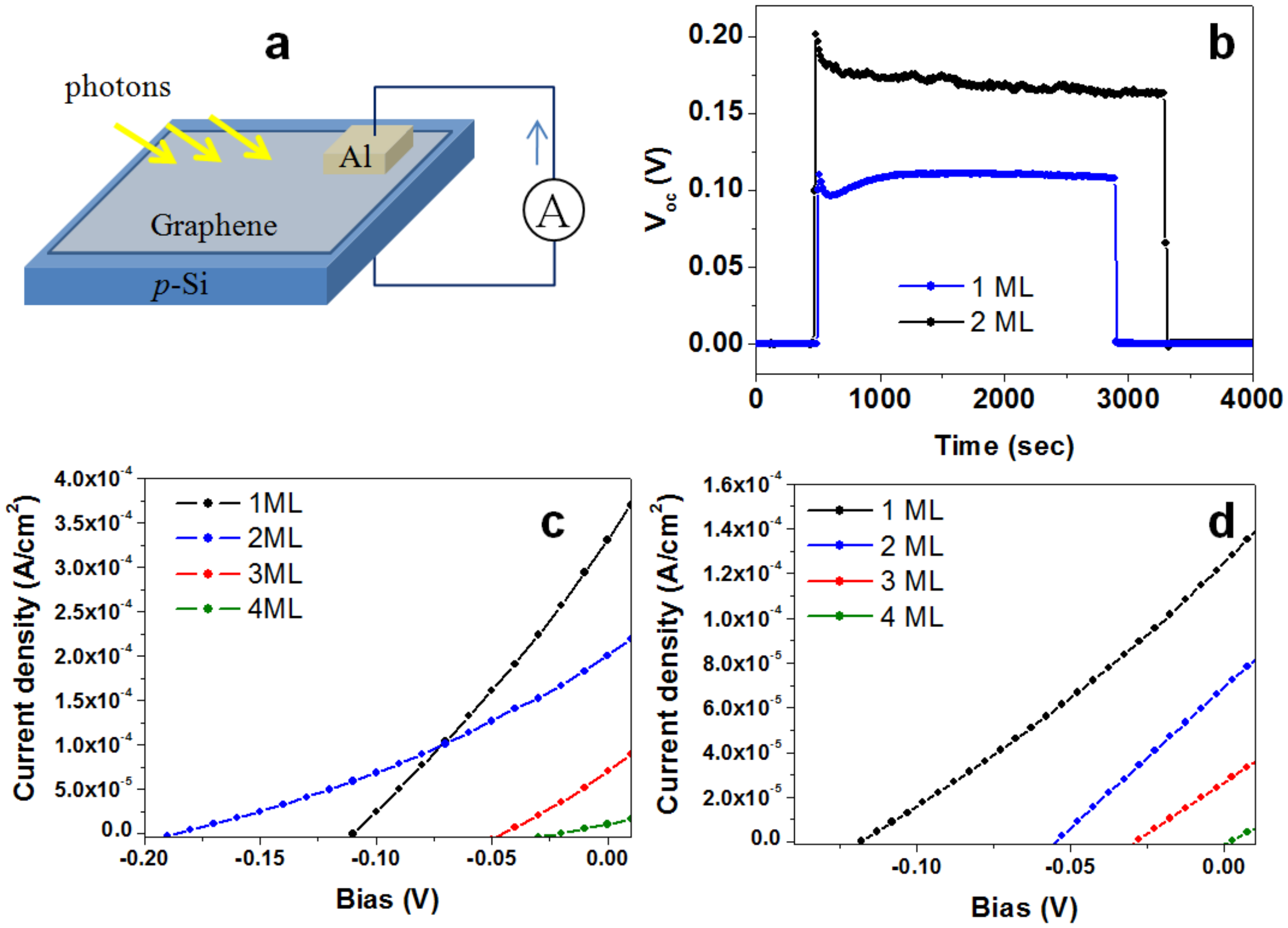

a
photons
Al
Graphene
p-Si
A
b
0.20
0.15
0.10
0.05
0.00
Voc (V)
1 ML
2 ML
0
1000
2000
3000
4000
Time (sec)
c
Current density (A/cm2)
4.0x10-4
3.5x10-4
3.0x10-4
2.5x10-4
2.0x10-4
1.5x10-4
1.0x10-4
5.0x10-5
0.0
1ML
2ML
3ML
4ML
-0.20
-0.15
-0.10
-0.05
0.00
Bias (V)
d
Current density (A/cm2)
1.6x10-4
1.4x10-4
1.2x10-4
1.0x10-4
8.0x10-5
6.0x10-5
4.0x10-5
2.0x10-5
0.0
1 ML
2 ML
3 ML
4 ML
-0.10
-0.05
0.00
Bias (V)

Figure 3.

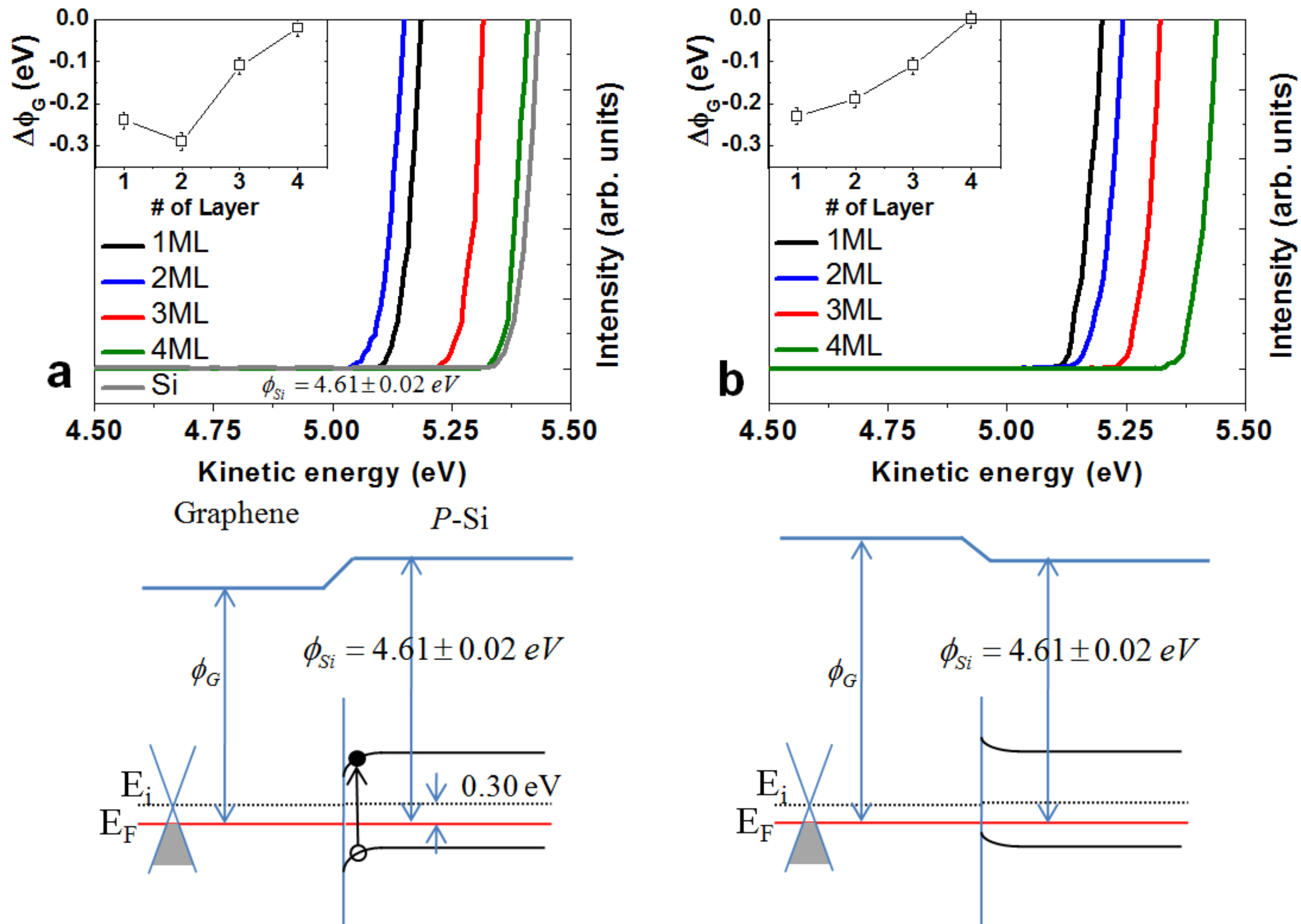

**c** Schottky contact, when $\phi_G < \phi_{si}$.

**d** Ohmic contact, when $\phi_G > \phi_{si}$.

Figure 4.

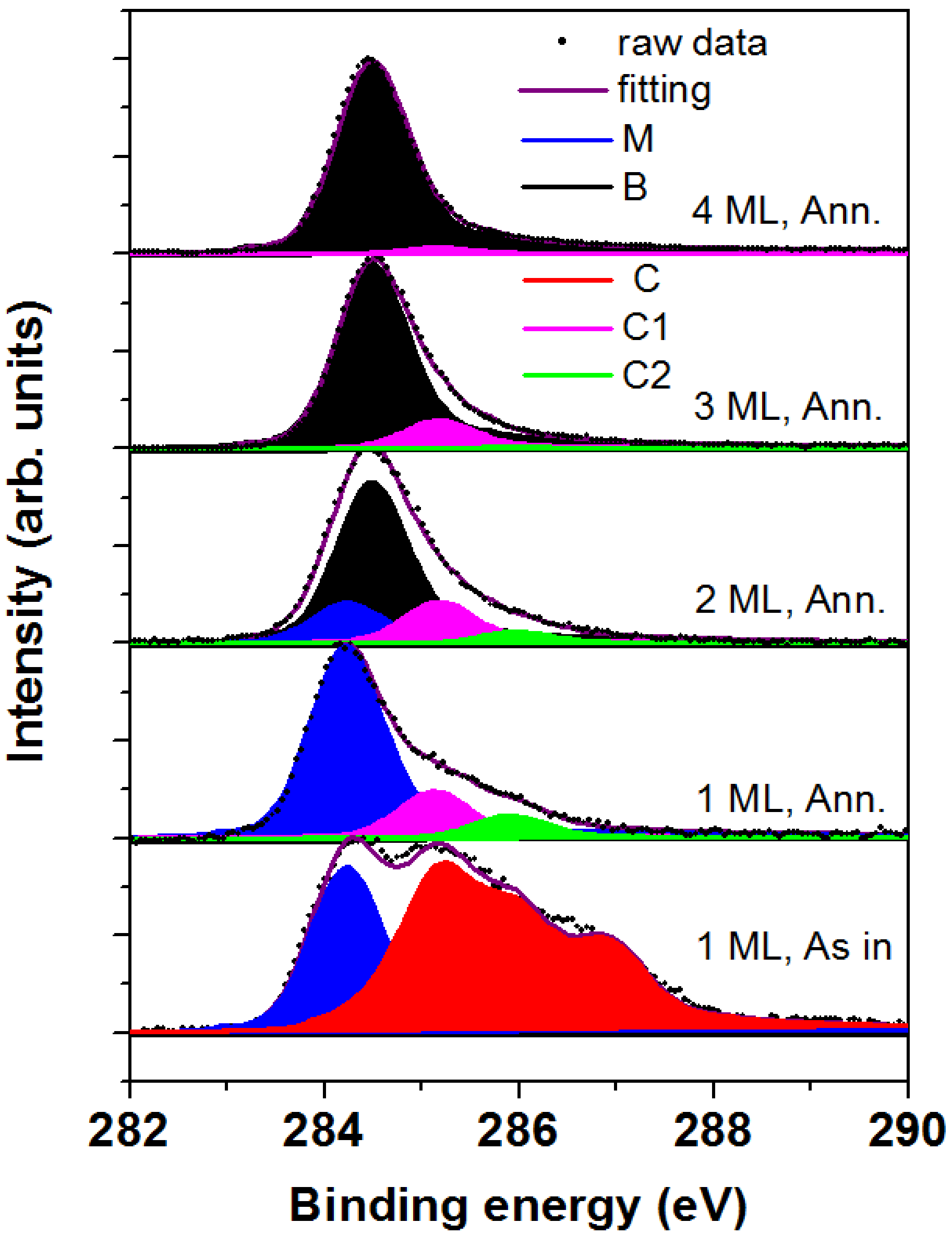

raw data
fitting
M
B
C
C1
C2
4 ML, Ann.
3 ML, Ann.
2 ML, Ann.
1 ML, Ann.
1 ML, As in
Intensity (arb. units)
282
284
286
288
290
Binding energy (eV)

Table S1.

| | # of Layer | $E_B$ (eV) | Area | L.W. (meV) | G.W. (meV) | ASY. |
|---|---|---|---|---|---|---|
| Annealed | 1 ML | 284.20 | 0.86 | 178 | 752 | 0.07 |
| | | 284.46 | 0.12 | 178 | 752 | 0.07 |
| | | 285.15 | 0.78 | 0 | 752 | 0.00 |
| | | 285.94 | 0.52 | 0 | 752 | 0.00 |
| | | 286.89 | 0.37 | 0 | 0 | 0.00 |
| As in. | 1 ML | 284.20 | 1.00 | 178 | 752 | 0.07 |
| | | 285.15 | 0.24 | 0 | 752 | 0.00 |
| | | 285.94 | 0.12 | 0 | 752 | 0.00 |
| | 2 ML | 284.20 | 0.21 | 178 | 752 | 0.07 |
| | | 284.46 | 0.84 | 178 | 752 | 0.07 |
| | | 285.15 | 0.22 | 0 | 752 | 0.00 |
| | | 285.94 | 0.06 | 0 | 752 | 0.00 |
| | 3 ML | 284.20 | 0.00 | 178 | 752 | 0.07 |
| | | 284.46 | 0.96 | 178 | 752 | 0.07 |
| | | 285.15 | 0.15 | 0 | 752 | 0.00 |
| | | 285.94 | 0.01 | 0 | 752 | 0.00 |
| | 4 ML | 284.20 | 0.00 | 178 | 752 | 0.07 |
| | | 284.46 | 1.00 | 178 | 752 | 0.07 |
| | | 285.15 | 0.03 | 0 | 752 | 0.00 |
| | | 285.94 | 0.01 | 0 | 752 | 0.00 |